\documentclass[conference]{IEEEtran}
\IEEEoverridecommandlockouts
\usepackage{cite}
\usepackage{amsmath,amssymb,amsfonts}
\usepackage{graphicx}
\usepackage{textcomp}
\usepackage{xcolor}
\usepackage{booktabs}
\usepackage{multirow}
\usepackage{array}
\usepackage{url}
\usepackage{balance}
\newcommand{\optionalfig}[4]{\begin{figure}[!t]\centering\IfFileExists{#1}{\includegraphics[width=0.98\linewidth]{#1}}{\fbox{\begin{minipage}[c][#2][c]{0.94\linewidth}\centering #3\end{minipage}}}\caption{#4}\end{figure}}
\newcommand{\optionalwidefig}[4]{\begin{figure*}[!t]\centering\IfFileExists{#1}{\includegraphics[width=0.96\textwidth]{#1}}{\fbox{\begin{minipage}[c][#2][c]{0.94\textwidth}\centering #3\end{minipage}}}\caption{#4}\end{figure*}}
\def\BibTeX{{\rm B\kern-.05em{\sc i\kern-.025em b}\kern-.08em
    T\kern-.1667em\lower.7ex\hbox{E}\kern-.125emX}}

\begin{document}

\title{TD-STGT: A Spatio-Temporal Graph Transformer for Mobile Traffic Demand Forecasting}

\author{\IEEEauthorblockN{Mohamad Alkadamani\IEEEauthorrefmark{1}\IEEEauthorrefmark{2} and Halim Yanikomeroglu\IEEEauthorrefmark{2}}
\IEEEauthorblockA{\IEEEauthorrefmark{1}Communications Research Centre Canada, Ottawa, ON, Canada}
\IEEEauthorblockA{\IEEEauthorrefmark{2}Department of Systems and Computer Engineering, Carleton University, Ottawa, ON, Canada}
\IEEEauthorblockA{Email: mohamad.alkadamani@ised-isde.gc.ca, halim@sce.carleton.ca}}

\maketitle

\begin{abstract}
Fine-grained mobile traffic demand forecasting is essential for long-term planning of 5G and future 6G networks, including radio upgrades, site densification, backhaul expansion, and spectrum activation. This paper proposes the Traffic Demand Spatio-Temporal Graph Transformer (TD-STGT), a graph neural forecasting framework for predicting changes in wireless mobile traffic demand across fine geographic grids. The framework uses a population-scaled demand proxy developed from crowdsourced mobile measurements and daytime population information. Experiments across five Canadian metropolitan regions show that TD-STGT achieves the best performance in forecasting grid-level demand changes, reaching a $\Delta R^2$ of 0.462 and reducing $\Delta$RMSE by 5.7\% relative to the strongest baseline. The proposed model provides a practical tool for identifying areas with increasing demand pressure and prioritizing future mobile-network capacity upgrades.
\end{abstract}


\section{Introduction}

Mobile network traffic demand continues to increase as LTE and 5G networks support high-resolution video, cloud services, fixed-wireless access, connected vehicles, and emerging 6G use cases \cite{ericsson2025}. For mobile network operators (MNOs), anticipating where demand will grow is essential for planning decisions such as radio upgrades, site densification, backhaul expansion, and spectrum activation. A key challenge is that direct operator traffic data are proprietary and are generally not accessible for research. This study therefore uses crowdsourced mobile measurement data as an observable basis for traffic-demand analysis. In this context, crowdsourced data refers to time-stamped and geo-located user-side measurements collected from mobile devices through software development kits (SDKs) embedded in applications. Such data do not provide the complete carried traffic of an operator network, but they provide repeated observations of mobile-network activity over space and time.

Prior spectrum-demand studies used geospatial features, deployment-based proxies, interpretable learning, and graph models to estimate local demand \cite{ParekhPimrc2023,ParekhOJCOMS2025,ParekhVTC2024DeepSpectrum,Alkadamani2024,AlkadamaniGlobecom2025}, but they mainly address spatial estimation. Mobile traffic forecasting has used recurrent and graph-based models to capture temporal and spatial dependence \cite{Qiu2018WirelessRNN,Wang2019Metropolis,Wang2023AHSTGNN,jiang2022gnn}; however, many studies rely on proprietary counters or short observation windows.

This paper proposes a data-driven framework that constructs a population-scaled proxy from crowdsourced measurements and forecasts grid-level demand change using a spatio-temporal graph transformer. Spatio-temporal modeling is used because demand does not evolve independently across grids: nearby and functionally related areas often change together. The proposed model, referred to as Traffic Demand Spatio-Temporal Graph Transformer (TD-STGT), uses local graph attention, global spatial self-attention, temporal convolution, and temporal self-attention to forecast grid-level demand changes.

The contributions are summarized as follows:
\begin{itemize}
    \item A crowdsourced proxy for mobile traffic demand is developed from user-side mobile measurement data.
    \item A spatio-temporal graph transformer architecture is formulated for grid-level mobile traffic-demand forecasting.
    \item An assessment is conducted across five Canadian metropolitan regions, demonstrating the ability of the proposed model to forecast demand-change patterns in urban and suburban planning scenarios.
\end{itemize}

The remainder of the paper is organized as follows. Section II presents the analysis setup and proxy construction. Section III formulates the forecasting problem and describes TD-STGT. Section IV presents the experimental setup and results. Section V concludes the paper.

\section{Analysis Setup and Proxy Construction}
\subsection{Analysis Setup}
The study covers Montreal, Ottawa, Toronto, Calgary, and Vancouver, focusing on urban and suburban areas where mobile traffic growth is concentrated. Each region is divided into grid cells of approximately $1.5\,\mathrm{km}^2$, yielding 38,246 grids. Crowdsourced observations are spatially assigned to grids and aggregated weekly and monthly; monthly aggregation is used as the main planning-level granularity because it is more stable under sampling variability.

The data include SDK-based crowdsourced measurements, daytime population, and network deployment records. The crowdsourced data provide user-side observations of mobile activity, daytime population approximates the number of potential users present in each grid during peak activity periods, and deployment records are used only for independent validation.

\subsection{Population-Scaled Peak-Demand Proxy}
The proposed proxy preserves user-side information in transmitted-byte measurements while reducing bias from uneven SDK sampling and low-sample grids. Let $\mathcal{M}_{g,t}^{p}$ be the set of crowdsourced observations in grid $g$ during peak hours of period $t$, and let $b_m$ be the transmitted bytes of observation $m$. The observed peak-hour byte volume and record count are
\begin{equation}
B^{p}_{g,t}=\sum_{m\in\mathcal{M}_{g,t}^{p}} b_m,
\quad
R^{p}_{g,t}=|\mathcal{M}_{g,t}^{p}|.
\label{eq:peakbytesrecords}
\end{equation}

The exposure-normalized peak intensity is
\begin{equation}
\beta_{g,t}^{p}=
\frac{B^{p}_{g,t}}{R^{p}_{g,t}+\epsilon},
\label{eq:bytesrecord}
\end{equation}
where $\epsilon$ avoids division by zero.

To stabilize low-sample grids, the grid-specific value is shrunk toward the corresponding city-period average:
\begin{equation}
\bar{\beta}_{c,t}^{p}=
\frac{\sum_{g\in c} B^{p}_{g,t}}
{\sum_{g\in c} R^{p}_{g,t}+\epsilon},
\quad
w_{g,t}=\frac{R^{p}_{g,t}}{R^{p}_{g,t}+\kappa},
\label{eq:shrinkage}
\end{equation}
where $c$ is the city containing grid $g$ and $\kappa$ controls shrinkage strength. The stabilized peak-hour intensity is
\begin{equation}
\tilde{\beta}_{g,t}^{p}
=
w_{g,t}\beta_{g,t}^{p}
+
(1-w_{g,t})\bar{\beta}_{c,t}^{p}.
\label{eq:smoothedbpr}
\end{equation}
When a grid has many peak-hour samples, $w_{g,t}$ approaches one; otherwise, the estimate is pulled toward the city-period average.

Let $P^{day}_{g}$ denote daytime population. The population-scaled proxy is
\begin{equation}
D^{pop}_{g,t}=P^{day}_{g}\tilde{\beta}_{g,t}^{p}.
\label{eq:popproxy}
\end{equation}
This proxy represents grid-level peak-demand potential by combining observed mobile data usage per crowdsourced observation with the daytime population in each grid. The level series is log-transformed as $y_{g,t}=\log(1+D^{pop}_{g,t})$, and the forecasting target is the one-period change $\Delta y_{g,t}=y_{g,t}-y_{g,t-1}$.
The proxy is computed at weekly and monthly resolution. Weekly aggregation provides more temporal samples, while monthly aggregation is used as the main planning-level target because it smooths sampling variability and aligns with MNO capacity-planning cycles.

\subsection{Proxy Behavior and Validation}
\begin{figure}[!t]
\centering
\IfFileExists{proxy_temporal_evolution.png}
{\includegraphics[width=0.92\columnwidth]{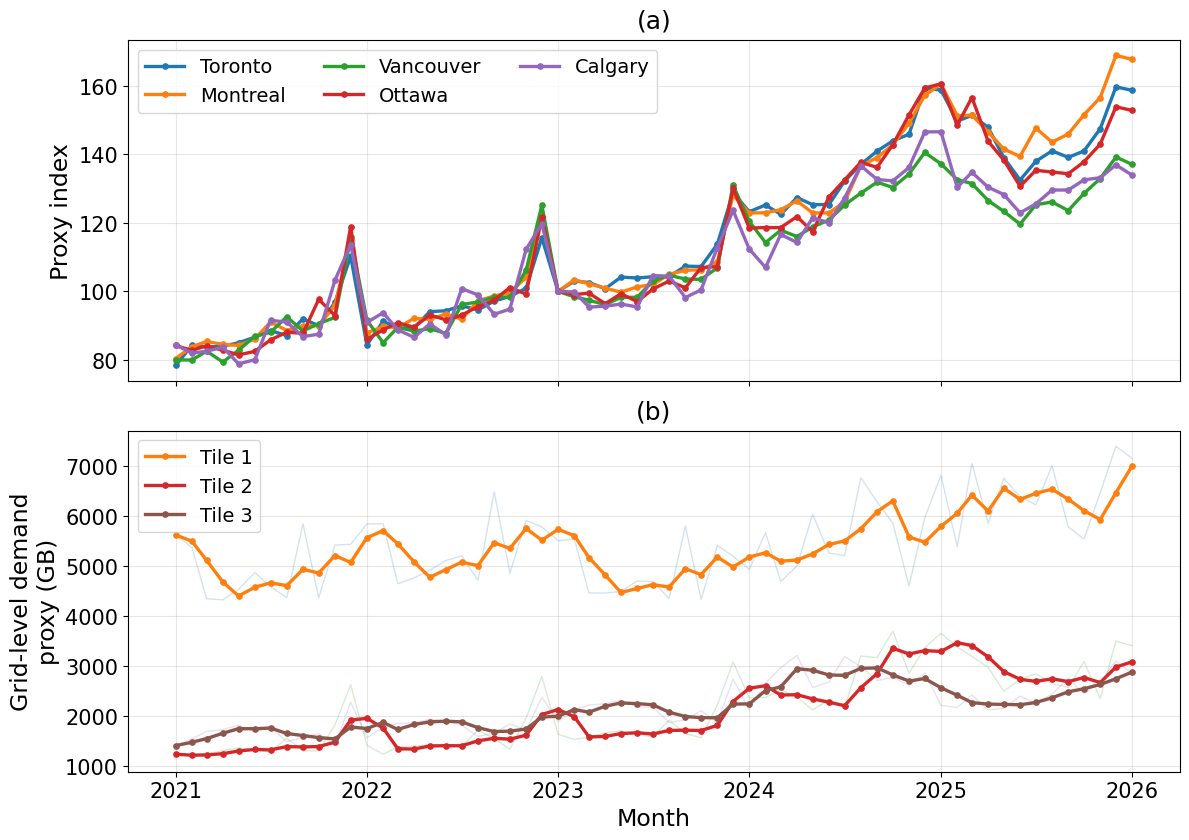}}
{\fbox{\begin{minipage}[c][1.15in][c]{0.94\columnwidth}
\centering Figure placeholder: insert temporal evolution plot of the proposed population-scaled peak-demand proxy.
\end{minipage}}}
\caption{Temporal evolution of the proposed population-scaled peak-demand proxy.}
\label{fig:proxy_temporal_evolution}
\end{figure}

Figure~\ref{fig:proxy_temporal_evolution} summarizes the temporal behavior of the proxy before forecasting. Panel (a) indexes the city-level aggregate series and highlights broad relative growth patterns across metropolitan regions. Panel (b) shows representative high-demand Toronto grids and illustrates that even within one city, local trajectories can evolve differently. This heterogeneity motivates grid-level spatio-temporal forecasting rather than a purely aggregate city-level predictor.

\begin{figure}[!t]
\centering
\IfFileExists{toronto_proxy_rate_change_map.png}
{\includegraphics[width=0.92\columnwidth]{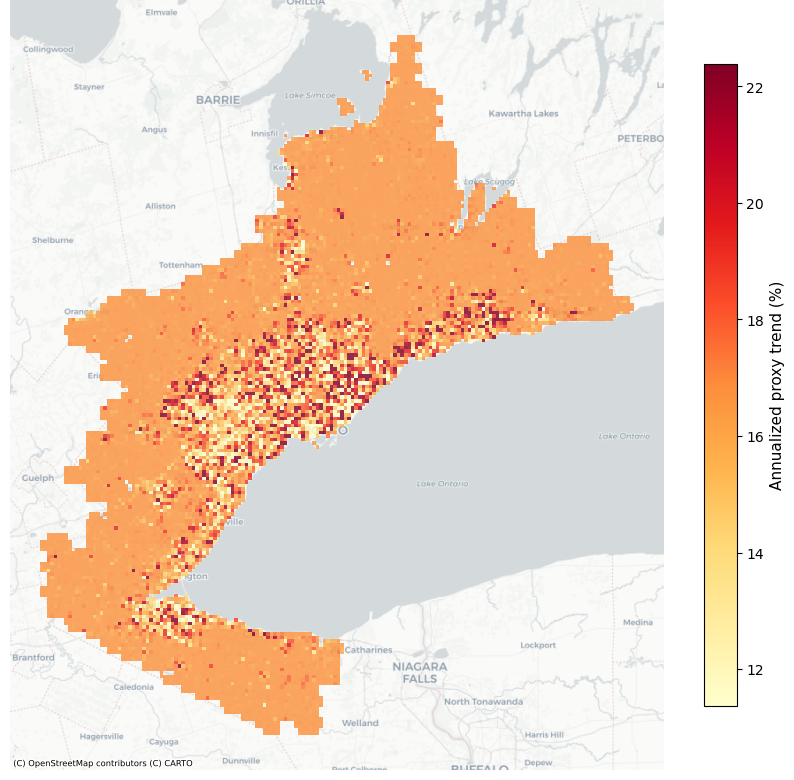}}
{\fbox{\begin{minipage}[c][1.15in][c]{0.94\columnwidth}
\centering Figure placeholder: insert Toronto spatial map of proxy rate-of-change by grid.
\end{minipage}}}
\caption{Spatial rate of change of the proposed proxy in Toronto.}
\label{fig:proxy_spatial_change}
\end{figure}

In Fig.~\ref{fig:proxy_spatial_change}, each grid is colored by the annualized trend of its monthly log-proxy series. The spatially heterogeneous growth pattern in Toronto further motivates graph-based forecasting.

Network deployment records are used for independent validation. A meaningful demand proxy should exhibit positive spatial association with deployed bandwidth, because operators tend to provision greater capacity in areas with persistent demand pressure. For validation, a site-data snapshot is assigned to the same grid. Let $BW_{g,t}$ denote total deployed bandwidth in grid $g$ \cite{Alkadamani2024}. The association is measured as
\begin{equation}
\rho_t(D^{pop},BW)=
\frac{\mathrm{Cov}_g(D^{pop}_{g,t},BW_{g,t})}
{\sigma_{D,t}\sigma_{BW,t}},
\label{eq:corr}
\end{equation}
where $\mathrm{Cov}_g(\cdot)$ is covariance across grids and $\sigma_{D,t}$ and $\sigma_{BW,t}$ are the corresponding standard deviations.

\section{Problem Formulation}
The objective is to forecast changes in grid-level mobile traffic demand represented by the crowdsourced population-scaled proxy. Let $\mathbf{y}_t\in\mathbb{R}^{N}$ denote the log-transformed proxy values for all $N$ grids at period $t$, where $y_{g,t}=\log(1+D^{pop}_{g,t})$. Because direct level forecasting is highly persistent, the main supervised target is the one-period change:
\begin{equation}
\Delta \mathbf{y}_t = \mathbf{y}_t-\mathbf{y}_{t-1}.
\label{eq:deltatarget}
\end{equation}
This formulation evaluates whether a model can identify grids whose demand trajectory changes beyond simple persistence. Let $\mathbf{X}_t\in\mathbb{R}^{N\times F}$ denote available period-level features, including historical proxy values, previous changes, lagged peak-hour sample counts, exposure-normalized byte indicators and daytime population descriptors. The grid-time panel contains $N=10{,}250$ grids across the five metropolitan regions.

For an input window of length $L$, the model receives historical proxy states, change values, and auxiliary grid-period features from $t-L$ to $t-1$ and predicts future proxy changes as
\begin{equation}
\widehat{\Delta\mathbf{Y}}_{t:t+H-1}=f_{\Theta}\left(\mathcal{X}_{t-L:t-1},\mathcal{G}\right),
\label{eq:forecasting}
\end{equation}
where $\mathcal{X}_{t-L:t-1}$ is the spatio-temporal input tensor, $\mathcal{G}$ is the grid graph, $H$ is the forecast horizon, and $\Theta$ denotes trainable parameters. The main task uses $H=1$, and evaluation reports $\Delta R^2$, $\Delta$RMSE, and $\Delta$MAE on the change target.

\section{Proposed TD-STGT Methodology}
\subsection{Graph Construction}
A spatial graph $\mathcal{G}=(\mathcal{V},\mathcal{E})$ is constructed where each node represents a grid cell. Edges encode spatial adjacency and distance-based proximity. The edge weight between nodes $i$ and $j$ is
\begin{equation}
A_{ij}=\begin{cases}
\exp\left(-\frac{\|s_i-s_j\|_2^2}{\sigma^2}\right), & j\in\mathcal{N}_k(i),\\
0, & \mathrm{otherwise},
\end{cases}
\label{eq:adj}
\end{equation}
where $s_i$ and $s_j$ are grid centroids, $\mathcal{N}_k(i)$ contains the $k$ nearest grids, and $\sigma$ controls distance decay. Self-loops preserve each grid's history; the default setting uses $k=8$.

\subsection{Overall Architecture}
\begin{figure*}[!t]
    \centering
    \includegraphics[width=0.90\textwidth]{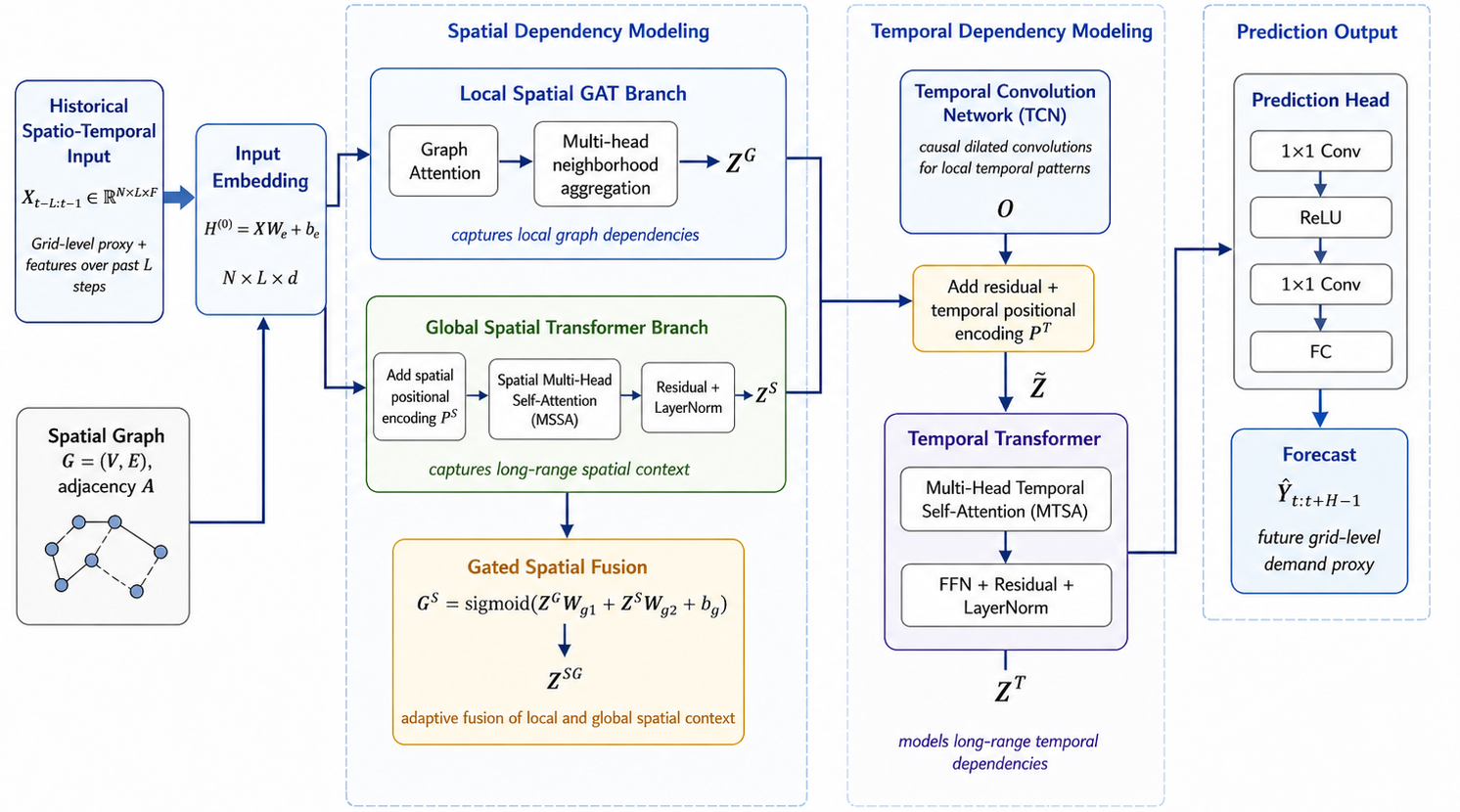}
    \caption{Overview of the TD-STGT architecture.}
    \label{fig:td_stgt_architecture}
\end{figure*}

TD-STGT contains spatial, temporal, and prediction modules. It combines local graph attention network (GAT), global spatial attention, temporal convolutional network (TCN), and temporal transformer blocks. Fig.~\ref{fig:td_stgt_architecture} summarizes the architecture. 

\subsection{Input Embedding}
The input window is projected into a $d$-dimensional representation:
\begin{equation}
\mathbf{H}^{(0)}=\phi_e(\mathcal{X}_{t-L:t-1})=\mathcal{X}_{t-L:t-1}\mathbf{W}_e+\mathbf{b}_e,
\end{equation}
where $\mathbf{H}^{(0)}\in\mathbb{R}^{N\times L\times d}$.

\subsection{Local Spatial Graph Attention Branch}
For each time step $\tau$, a graph attention layer updates node $i$ using its spatial neighbors:
\begin{equation}
e_{ij}^{\tau}=\mathrm{LeakyReLU}\left(\mathbf{a}^{\top}[\mathbf{W}_G\mathbf{h}_{i,\tau}\Vert \mathbf{W}_G\mathbf{h}_{j,\tau}]\right),
\end{equation}
\begin{equation}
\alpha_{ij}^{\tau}=\frac{\exp(e_{ij}^{\tau})}{\sum_{k\in\mathcal{N}(i)}\exp(e_{ik}^{\tau})},
\end{equation}
\begin{equation}
\mathbf{z}^{G}_{i,\tau}=\sigma\left(\sum_{j\in\mathcal{N}(i)}\alpha_{ij}^{\tau}\mathbf{W}_G\mathbf{h}_{j,\tau}\right),
\label{eq:gat}
\end{equation}
where $e_{ij}^{\tau}$ is the unnormalized attention score from neighbor $j$ to grid $i$, $\alpha_{ij}^{\tau}$ is the normalized attention coefficient, $\mathbf{a}$ is a learnable attention vector, $\mathbf{W}_G$ is a graph-attention projection matrix, $\mathbf{h}_{i,\tau}$ is the input embedding of grid $i$ at time $\tau$, $\Vert$ denotes concatenation, and $\sigma(\cdot)$ is a nonlinear activation. Multi-head attention is implemented by concatenating or averaging head-specific outputs.

\subsection{Global Spatial Transformer Branch}
The spatial transformer captures long-range relationships using self-attention. Spatial positional encodings are added as
\begin{equation}
\tilde{\mathbf{H}}_{:,\tau,:}=\mathbf{H}^{(0)}_{:,\tau,:}+\mathbf{P}^{S}.
\end{equation}
For each time step, multi-head spatial self-attention is computed as
\begin{equation}
\mathrm{MSSA}(\tilde{\mathbf{H}})=\mathrm{Concat}(\mathrm{head}_1,\ldots,\mathrm{head}_h)\mathbf{W}^{O},
\end{equation}
\begin{equation}
\mathrm{head}_m=\mathrm{softmax}\left(\frac{\mathbf{Q}_m^S(\mathbf{K}_m^S)^{\top}}{\sqrt{d_h}}\right)\mathbf{V}_m^S,
\end{equation}
where $h$ is the number of heads, $d_h$ is the per-head dimension, and $\mathbf{Q}_m^S$, $\mathbf{K}_m^S$, and $\mathbf{V}_m^S$ are query, key, and value matrices, respectively. The branch output is
\begin{equation}
\mathbf{Z}^{S}=\mathrm{LN}\left(\mathrm{MSSA}(\tilde{\mathbf{H}})+\tilde{\mathbf{H}}\right),
\end{equation}
where $\mathrm{LN}(\cdot)$ denotes layer normalization.

\subsection{Gated Spatial Fusion}
The local GAT output $\mathbf{Z}^{G}$ and global spatial transformer output $\mathbf{Z}^{S}$ are fused through a nonlinear gate:
\begin{equation}
\mathbf{G}^{S}=\mathrm{sigmoid}\left(\mathbf{Z}^{G}\mathbf{W}_{g1}+\mathbf{Z}^{S}\mathbf{W}_{g2}+\mathbf{b}_g\right),
\end{equation}
\begin{equation}
\mathbf{Z}^{SG}=\mathbf{G}^{S}\odot \mathbf{Z}^{S}+(1-\mathbf{G}^{S})\odot \mathbf{Z}^{G},
\label{eq:gate}
\end{equation}
where $\mathbf{G}^{S}$ is the learned fusion gate, $\mathbf{W}_{g1}$ and $\mathbf{W}_{g2}$ are trainable projection matrices, $\mathbf{b}_g$ is a bias term, and $\odot$ denotes element-wise multiplication. The gate allows each grid and timestamp to determine whether local or global spatial context is more informative.

\subsection{Temporal Dependency Modeling}
The temporal module begins with a TCN applied to each grid sequence, using causal dilated convolutions to model local temporal dependencies. A causal dilated convolution is
\begin{equation}
\mathbf{o}_{i,\tau}=\sum_{q=0}^{K-1}\mathbf{W}_{q}\mathbf{Z}^{SG}_{i,\tau-q\cdot d_r},
\end{equation}
where $\mathbf{o}_{i,\tau}$ is the TCN output for grid $i$ at time $\tau$, $K$ is the convolution kernel size, $d_r$ is the dilation rate, and $\mathbf{W}_q$ is the convolution weight at lag $q$. Residual TCN blocks capture local temporal changes in deployment and demand.

Temporal positional encodings are then added:
\begin{equation}
\tilde{\mathbf{Z}}=\mathbf{O}+\mathbf{Z}^{SG}+\mathbf{P}^{T},
\end{equation}
where $\mathbf{O}$ is the TCN representation and $\mathbf{P}^{T}$ encodes the order of the input periods. Temporal multi-head self-attention is then applied along the time dimension for each grid:
\begin{equation}
\mathrm{MTSA}(\tilde{\mathbf{Z}})=\mathrm{Concat}(\mathrm{thead}_1,\ldots,\mathrm{thead}_h)\mathbf{W}^{T,O},
\end{equation}
\begin{equation}
\mathrm{thead}_m=\mathrm{softmax}\left(\frac{\mathbf{Q}_m^T(\mathbf{K}_m^T)^{\top}}{\sqrt{d_h}}\right)\mathbf{V}_m^T,
\end{equation}
where $\mathbf{Q}_m^T$, $\mathbf{K}_m^T$, and $\mathbf{V}_m^T$ are temporal query, key, and value matrices for head $m$, and $\mathbf{W}^{T,O}$ is the temporal output projection. The temporal representation is
\begin{equation}
\mathbf{Z}^{T}=\mathrm{LN}\left(\mathrm{FFN}\left(\mathrm{LN}(\mathrm{MTSA}(\tilde{\mathbf{Z}})+\tilde{\mathbf{Z}})\right)+\tilde{\mathbf{Z}}\right),
\end{equation}
where $\mathrm{FFN}(\cdot)$ is a position-wise feed-forward network.

\subsection{Prediction Head and Loss Function}
The prediction head maps the temporal representation to the forecast horizon of proxy changes:
\begin{equation}
\widehat{\Delta\mathbf{Y}}_{t:t+H-1}
=
\phi_{\mathrm{out}}(\mathbf{Z}^{T}),
\label{eq:predhead}
\end{equation}
where $\phi_{\mathrm{out}}(\cdot)$ is implemented using two $1\times1$ convolutional layers with a ReLU activation followed by a fully connected projection to the forecast horizon.

The model is trained using a loss function that combines change-forecasting error, spatial smoothness, and temporal smoothness:
\begin{equation}
\mathcal{L}=\mathcal{L}_{\Delta}+\lambda_s\mathcal{L}_{S}+\lambda_t\mathcal{L}_{T},
\end{equation}
where $\lambda_s$ and $\lambda_t$ control spatial and temporal regularization:
\begin{equation}
\mathcal{L}_{\Delta}=\frac{1}{NH}\sum_{g=1}^{N}\sum_{h=0}^{H-1}\left(\Delta y_{g,t+h}-\widehat{\Delta y}_{g,t+h}\right)^2,
\end{equation}
\begin{equation}
\mathcal{L}_{S}=\frac{1}{|\mathcal{E}|H}\sum_{(i,j)\in\mathcal{E}}\sum_{h=0}^{H-1}\left(\widehat{\Delta y}_{i,t+h}-\widehat{\Delta y}_{j,t+h}\right)^2, \text{and}
\end{equation}
\begin{equation}
\mathcal{L}_{T}=\frac{1}{N(H-1)}\sum_{g=1}^{N}\sum_{h=1}^{H-1}\left(\widehat{\Delta y}_{g,t+h}-\widehat{\Delta y}_{g,t+h-1}\right)^2,
\end{equation}
where $\mathcal{L}_{\Delta}$ penalizes change-forecasting error, $\mathcal{L}_{S}$ penalizes unrealistic discontinuities in predicted changes between connected grids, and $\mathcal{L}_{T}$ regularizes abrupt oscillations across multi-step change forecasts.

\section{Experimental Setup, Results, and Discussion}

\subsection{Dataset Split and Sliding Window Strategy}
Chronological splitting is used to avoid future information leakage: earlier periods are used for training, an intermediate period for validation, and the most recent periods for testing. Validation and test samples use only historical context from preceding periods, so no future proxy value or current-period proxy component enters the input.

To assess spatial generalization, an additional blocked split holds out contiguous grid regions during training. The strictest protocol combines both dimensions by testing future periods in spatially held-out regions. The main results report the chronological test setting, while the spatial and spatio-temporal splits are used for robustness analysis.

The main configuration uses an input window of approximately four years and $H=1$ forecast step, corresponding to next-period proxy-change forecasting at the selected weekly or monthly granularity.

\subsection{Model Configuration}
Table \ref{tab:config} lists the default TD-STGT configuration. All models are implemented in PyTorch 2.7.1 and trained on an AWS EC2 G5.2xlarge instance equipped with one NVIDIA A10G GPU with 24 GB GPU memory. Hyperparameters are selected through validation-set grid search over reasonable ranges commonly used for spatio-temporal forecasting.

\begin{table}[!t]
\centering
\caption{TD-STGT model configuration}
\label{tab:config}
\footnotesize
\begin{tabular}{ll}
\toprule
Parameter & Value \\
\midrule
Grid nodes $N$ & 38,246 \\
Input features $F$ & 18 proxy-derived and auxiliary features \\
Input window $L$ & 48 snapshots \\
Forecast horizon $H$ & 1 snapshot \\
Graph neighbors $k$ & 8 \\
Embedding dimension $d$ & 128 \\
Attention heads $h$ & 4 \\
TCN kernel size $K$ & 3 \\
Dropout & 0.20 \\
Optimizer & AdamW \\
Learning rate & $1\times10^{-3}$ \\
Batch size & 32 windows \\
Maximum epochs & 100 \\
Early stopping & 12 epochs \\
AWS instance & EC2 G5.2xlarge \\
GPU & NVIDIA A10G 24 GB \\
\bottomrule
\end{tabular}
\end{table}

\subsection{Baseline Models and Evaluation Metrics}
Seven baseline models are selected: linear autoregressive regression, RF, XGBoost, LSTM, GRU, CNN-LSTM, and STGCN. The linear model provides an interpretable time-series baseline. RF and XGBoost evaluate strong nonlinear tabular approaches. LSTM and GRU evaluate sequence models without explicit spatial graph structure. CNN-LSTM evaluates Euclidean spatio-temporal learning. STGCN evaluates graph-based spatio-temporal forecasting. Table~\ref{tab:baselinehp} summarizes the main hyperparameters used for the baseline models.

\begin{table}[!t]
\centering
\caption{Key hyperparameters of baseline models}
\label{tab:baselinehp}
\begin{tabular}{lll}
\toprule
Model & Hyperparameter & Value \\
\midrule
Linear ARX & Regularization & Ridge, $\alpha=1.0$ \\
RF & Estimators, depth & 300, 18 \\
XGBoost & Estimators, LR, depth & 500, 0.03, 6 \\
LSTM & Layers, hidden, dropout & 2, 96, 0.2 \\
GRU & Layers, hidden, dropout & 2, 96, 0.2 \\
CNN-LSTM & Kernel, hidden, dropout & 3, 96, 0.2 \\
STGCN & Layers, channels, kernel & 2, 64, 3 \\
TD-STGT & Heads, embedding, dropout & 4, 128, 0.2 \\
\bottomrule
\end{tabular}
\end{table}

Performance is evaluated with $\Delta R^2$, $\Delta$RMSE, and $\Delta$MAE on the one-period change target. These metrics avoid overstating skill from the high persistence of proxy levels.

\subsection{Proxy Validation Results}
Table \ref{tab:proxy} reports deployed-bandwidth validation results for alternative proxy variants. The proposed population-scaled peak-demand proxy reaches the strongest association with deployed bandwidth, with median Pearson correlation of 0.71 and OLS $R^2$ of 0.66.

\begin{table}[!t]
\centering
\caption{Proxy validation against deployed bandwidth}
\label{tab:proxy}
\scriptsize
\begin{tabular}{p{0.48\linewidth}ccc}
\toprule
Proxy variant & Median $\rho$ & OLS $R^2$ & $p$-value \\
\midrule
Observed all-day bytes/hour & 0.64 & 0.60 & $<0.001$ \\
Population-scaled peak-demand proxy & \textbf{0.71} & \textbf{0.66} & $<0.001$ \\
\bottomrule
\end{tabular}
\end{table}

\subsection{Forecasting Performance}
Table \ref{tab:mainresults} compares TD-STGT against the persistence and learning baselines for one-step-ahead change forecasting. The no-change baseline predicts $\Delta y_{g,t}=0$ and is included because the proxy level is highly persistent. TD-STGT achieves the highest $\Delta R^2$ and the lowest $\Delta$RMSE and $\Delta$MAE. The strongest non-proposed baseline is STGCN with $\Delta$RMSE of 0.053, while TD-STGT reduces $\Delta$RMSE to 0.050, a 5.7\% reduction.

\begin{table}[!t]
\centering
\caption{One-step demand-change forecasting performance on the 2025 test set}
\footnotesize
\label{tab:mainresults}
\begin{tabular}{lccc}
\toprule
Model & $\Delta R^2\uparrow$ & $\Delta$RMSE$\downarrow$ & $\Delta$MAE$\downarrow$ \\
\midrule
No-change baseline & 0.000 & 0.069 & 0.047 \\
Linear ARX & 0.112 & 0.065 & 0.045 \\
RF & 0.246 & 0.060 & 0.041 \\
XGBoost & 0.312 & 0.057 & 0.039 \\
LSTM & 0.274 & 0.059 & 0.040 \\
GRU & 0.291 & 0.058 & 0.039 \\
CNN-LSTM & 0.341 & 0.056 & 0.038 \\
STGCN & 0.405 & 0.053 & 0.036 \\
TD-STGT (Proposed) & \textbf{0.462} & \textbf{0.050} & \textbf{0.034} \\
\bottomrule
\end{tabular}
\end{table}

The no-change baseline captures temporal persistence, while RF, XGBoost, and recurrent models provide moderate gains. STGCN is the strongest baseline because it uses graph structure, and TD-STGT performs best by combining graph locality, global spatial context, and temporal attention.

\begin{figure}[!t]
\centering
\IfFileExists{absolute_error_distributions.png}
{\includegraphics[width=0.92\columnwidth]{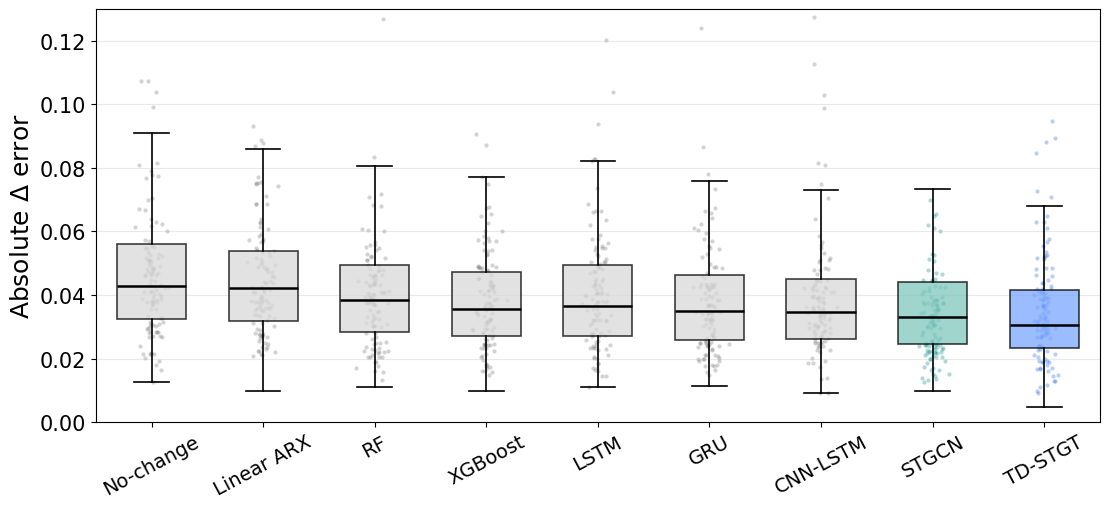}}
{\fbox{\begin{minipage}[c][1.05in][c]{0.94\columnwidth}
\centering Figure placeholder: insert absolute delta-error box plots by model.
\end{minipage}}}
\caption{Absolute delta-error distributions across models.}
\label{fig:box}
\end{figure}

Fig.~\ref{fig:box} shows the distribution of absolute delta errors. TD-STGT has the lowest median error and one of the tightest spreads, while STGCN is the strongest baseline and the recurrent models show larger dispersion.

\begin{figure}[!t]
\centering
\IfFileExists{robustness_random_seeds.png}
{\includegraphics[width=0.92\columnwidth]{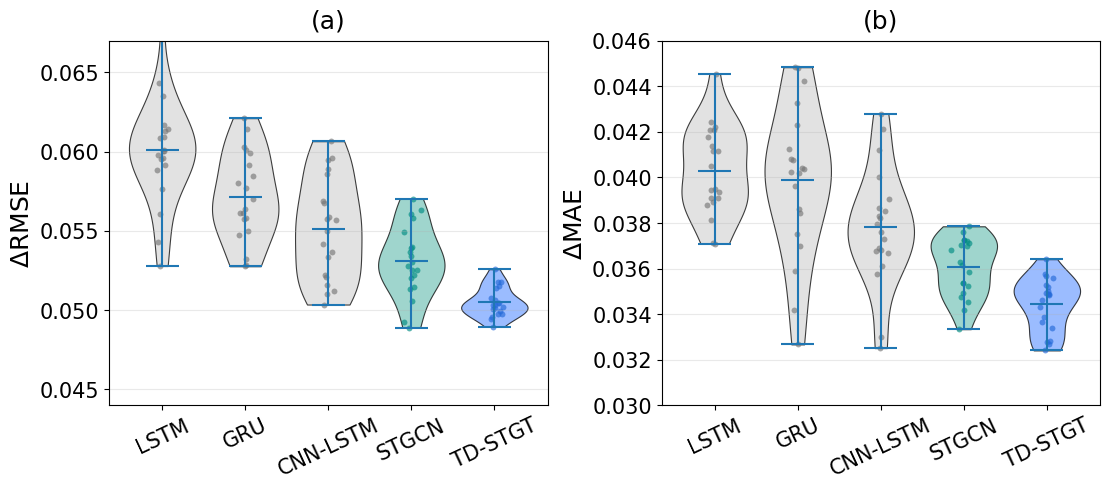}}
{\fbox{\begin{minipage}[c][1.05in][c]{0.94\columnwidth}
\centering Figure placeholder: insert $\Delta$RMSE and $\Delta$MAE violin plots across 20 random seeds.
\end{minipage}}}
\caption{Change-forecasting robustness across 20 random seeds.}
\label{fig:violin}
\end{figure}

The robustness analysis trains each neural model under 20 random seeds. Fig.~\ref{fig:violin} shows the resulting $\Delta$RMSE and $\Delta$MAE distributions. LSTM, GRU, and CNN-LSTM exhibit wider spreads, STGCN is more stable, and TD-STGT has the narrowest distribution, indicating the most consistent convergence.

\subsection{Sensitivity Analysis}
Sensitivity is evaluated by varying learning rate, attention heads, batch size, and embedding dimension.

\begin{table}[!t]
\centering
\caption{Sensitivity analysis for key hyperparameters}
\footnotesize
\label{tab:sensitivity}
\begin{tabular}{lccc}
\toprule
Hyperparameter & Setting & $\Delta$RMSE & $\Delta$MAE \\
\midrule
\multirow{3}{*}{Learning rate} & $5\times10^{-4}$ & 0.053 & 0.036 \\
 & $1\times10^{-3}$ & \textbf{0.050} & \textbf{0.034} \\
 & $2\times10^{-3}$ & 0.054 & 0.037 \\
\midrule
\multirow{3}{*}{Attention heads} & 2 & 0.054 & 0.037 \\
 & 4 & \textbf{0.050} & \textbf{0.034} \\
 & 8 & 0.052 & 0.035 \\
\midrule
\multirow{3}{*}{Batch size} & 16 & 0.053 & 0.036 \\
 & 32 & \textbf{0.050} & \textbf{0.034} \\
 & 64 & 0.052 & 0.035 \\
\midrule
\multirow{3}{*}{Embedding dimension} & 64 & 0.056 & 0.038 \\
 & 128 & \textbf{0.050} & \textbf{0.034} \\
 & 256 & 0.050 & 0.034 \\
\bottomrule
\end{tabular}
\end{table}

As shown in Table \ref{tab:sensitivity}, the best operating point uses learning rate $10^{-3}$, four heads, batch size 32, and embedding dimension 128.

\subsection{Ablation Study}
The ablation study removes one TD-STGT component at a time under the same training protocol to identify where the gains originate. As shown in Table \ref{tab:ablation}, removing the spatial module causes the largest degradation, confirming the importance of spatial context; proxy simplification also reduces accuracy.

\begin{table}[!t]
\centering
\caption{Ablation analysis of TD-STGT}
\label{tab:ablation}
\footnotesize
\begin{tabular}{p{0.48\linewidth}ccc}
\toprule
Variant & $\Delta R^2\uparrow$ & $\Delta$RMSE$\downarrow$ & $\Delta$MAE$\downarrow$ \\
\midrule
Full TD-STGT & \textbf{0.462} & \textbf{0.050} & \textbf{0.034} \\
Without spatial module & 0.311 & 0.063 & 0.043 \\
Without local GAT branch & 0.421 & 0.052 & 0.036 \\
Without spatial transformer & 0.394 & 0.054 & 0.037 \\
Without TCN & 0.372 & 0.055 & 0.038 \\
Without temporal transformer & 0.386 & 0.054 & 0.037 \\
Without user-weighted proxy mapping & 0.335 & 0.061 & 0.041 \\
\bottomrule
\end{tabular}
\end{table}

\subsection{Model Complexity and Computational Cost}
Table \ref{tab:complexity} compares advanced models in terms of parameter count, training time, inference time, and $\Delta$RMSE. TD-STGT is the most expensive model, but the cost remains practical for offline planning and is justified by the accuracy gain.

\begin{table}[!t]
\centering
\caption{Model complexity and computational cost}
\footnotesize
\label{tab:complexity}
\begin{tabular}{lcccc}
\toprule
Model & Params & Train Time & Inference & $\Delta$RMSE \\
 & (M) & (min) & (ms/batch) & \\
\midrule
LSTM & 0.31 & 38 & 1.7 & 0.059 \\
GRU & 0.27 & 36 & 1.5 & 0.058 \\
CNN-LSTM & 0.58 & 47 & 2.2 & 0.056 \\
STGCN & 1.12 & 59 & 2.9 & 0.053 \\
TD-STGT & 2.86 & 88 & 4.4 & \textbf{0.050} \\
\bottomrule
\end{tabular}
\end{table}

\section{Conclusion}
This study presented a spatio-temporal framework for forecasting grid-level changes in mobile traffic demand using a crowdsourced population-scaled proxy. The framework converts user-side mobile measurements and daytime population information into a demand signal whose temporal changes can be modeled at fine spatial granularity. Pilot results show that TD-STGT achieves the best performance among the evaluated baselines, with $\Delta R^2=0.462$ and $\Delta$RMSE of 0.050. The results indicate that proxy-driven change forecasting can help identify areas with increasing demand pressure and support proactive planning decisions such as densification, carrier activation, radio upgrades, and backhaul expansion.

\enlargethispage{20\baselineskip}
\balance
\bibliographystyle{IEEEtran}
\bibliography{references}

\end{document}